\documentclass[english,a4paper,american]{article}
\usepackage[T1]{fontenc}
\usepackage[latin1]{inputenc}
\usepackage{amsmath}
\usepackage{graphicx}
\usepackage{amssymb}

\makeatletter



\usepackage{babel}
\makeatother
\begin{document}

\title{Neutrino emission due to Cooper pairing in neutron stars}

\author{L. B. Leinson$^{1,2}$ and A. P\'{e}rez$²$ \\
 $^{1}$Institute of Terrestrial Magnetism, Ionosphere and\\
 Radio Wave Propagation\\
 RAS, 142190 Troitsk, Moscow Region, Russia\\
 $^{2}$Departamento de F\'{\i}sica Te\'{o}rica and IFIC, \\
 Universidad de Valencia-CSIC, \\
 Dr. Moliner 50, 46100--Burjassot, Valencia, Spain}

\maketitle
\begin{abstract}
Neutrino emission caused by Cooper pairing of baryons in neutron stars
is recalculated by accurately taking into account for conservation
of the vector weak current. The vector current contribution to the
neutrino emissivity is found to be several orders of magnitude smaller
than that obtained before by different authors. Therefore, the neutrino
energy losses due to singlet-state pairing of baryons can in practice
be neglected in simulations of neutron star cooling. This makes negligible
the neutrino radiation from pairing of protons or hyperons. The neutrino
radiation from triplet pairing takes place through axial weak currents.
For these states, when the total momentum projection is $m_{j}=0$,
the vanishing of the vector weak current contribution results in the
suppression of the neutrino energy losses by about 25\%. The neutrino
emissivity due to triplet pairing with $\left\vert m_{j}\right\vert =2$
is suppressed by about a factor of 3, caused by the collective contribution
of spin-density fluctuations in the condensate. 
\end{abstract}

\section{Introduction}

Thermal excitations in superfluid baryon matter of neutron stars,
in the form of broken Cooper pairs, can recombine into the condensate
by emitting neutrino pairs via neutral weak currents. The neutrino
energy losses caused by the pairing of neutrons in a singlet state
was first calculated by Flowers et al. \cite{FRS76} and reproduced
by other authors \cite{Vosk}, \cite{YKL98}. Yakovlev et al. \cite{YKL98}
have also considered the neutrino energy losses in the case of neutron
pairing in a triplet state. Jaikumar \& Prakash \cite{prak} have
generalized this mechanism to superconducting quark matter. The neutrino
energy losses due to pairing of hyperons \cite{Balberg}, \cite{Schaab}
are also discussed in the literature as possible cooling mechanisms
for superdense baryonic matter in neutron stars. Nowadays, these ideas
are widely accepted and used in numerical simulations of neutron star
evolution \cite{Sch}, \cite{Page98}, \cite{Yak98}. Nevertheless,
the existing theory of neutrino radiation from Cooper pairing of baryons
(quarks) in neutron stars rises some questions because, as we show
below, it is inconsistent with the hypothesis of conservation of the
vector current in weak interactions.

Let us recall shortly the main steps in the above calculations. The
low-energy Lagrangian of the weak interaction may be described by
a point-like current-current approach. For interactions mediated by
neutral weak currents, it can be written as%
\footnote{In what follows we use the Standard Model of weak interactions, the
system of units $\hbar=c=1$ and the Boltzmann constant $k_{B}=1$.
The fine-structure constant is $\alpha=e^{2}/4\pi=1/137$.%
} \begin{equation}
\mathcal{L}_{vac}=\frac{G_{F}}{2\sqrt{2}}J_{B}^{\mu}l_{\mu}.\end{equation}
 Here $G_{F}$ is the Fermi coupling constant, and the neutrino weak
current is given by $l_{\mu}=\bar{\nu}\gamma_{\mu}\left(1-\gamma_{5}\right)\nu$.
The vacuum weak current of the fermion (baryon or quark) is of the
form $J_{\mu}=\bar{\psi}\left(C_{V}\gamma_{\mu}-C_{A}\gamma_{\mu}\gamma_{5}\right)\psi$,
where, $\psi$ represents the fermion field, and the weak vertex includes
the vector and axial-vector terms with the corresponding coupling
constants $C_{V}$ and $C_{A}$.

Since relativistic calculations are more complicated and less transparent,
we consider the nonrelativistic case, typical for superfluid baryon
matter in neutron stars. Then, the nonrelativistic limits for the
baryon operators are $\bar{\psi}_{B}\gamma^{0}\psi_{B}\rightarrow\hat{\Psi}_{B}^{+}\hat{\Psi}_{B}$,
$\bar{\psi}_{B}\gamma_{i}\gamma_{5}\psi_{B}\rightarrow\hat{\Psi}_{B}^{+}\hat{\sigma}_{i}\hat{\Psi}_{B}$,
all others being zero. Here $\hat{\Psi}_{B}$ is the second-quantized
nonrelativistic spinor wave function, and $\hat{\sigma}_{i}$ are
the Pauli matrices.

The process is kinematically allowed due to the existence of a superfluid
energy gap $\Delta$, which admits the transition with time-like momentum
transfer $K=\left(\omega,\mathbf{k}\right)$, as required by the final
neutrino pair. We have $\omega>2\Delta$ and $\omega>k$ . The emissivity
for neutrino pairs due to recombination of baryon quasi-particle excitations
in the singlet state is then%
\footnote{In the case of $^{1}S_{0}$ pairing, the total spin of the Cooper
pair is zero, and the axial contribution vanishes in the nonrelativistic
limit.%
}\begin{eqnarray}
Q & = & \left(\frac{G_{F}}{2\sqrt{2}}\right)^{2}C_{V}^{2}\int\frac{d^{3}pd^{3}p^{\prime}}{\left(2\pi\right)^{8}}f\left(\epsilon_{\mathbf{p}}\right)f\left(\epsilon_{\mathbf{p}^{\prime}}\right)\int\frac{d^{3}k_{1}}{2\omega_{1}}\frac{d^{3}k_{2}}{2\omega_{2}}\omega\left\vert \mathcal{M}_{B}\right\vert ^{2}\left\vert \mathcal{M}_{\nu}\right\vert ^{2}\notag\\
 &  & \times\delta\left(\mathbf{p+p}^{\prime}-\mathbf{k}\right)\delta\left(\epsilon_{\mathbf{p}}+\epsilon_{\mathbf{p}^{\prime}}-\omega\right).\label{Gold}\end{eqnarray}
 Here $\omega=\omega_{1}+\omega_{2}$ and $\mathbf{k=k}_{1}+\mathbf{k}_{2}$
are the energy and momentum carried out by the freely escaping neutrino
pair, and $f\left(\epsilon_{p}\right)$ is the Fermi distribution
function of the quasi-particles with energy $\epsilon_{p}$, as given
by Eq. (\ref{p0}). The neutrino matrix element is of the standard
form $\left\vert \mathcal{M}_{\nu}\right\vert ^{2}=8\left(\omega_{1}\omega_{2}-\mathbf{k}_{1}\mathbf{k}_{2}\right)$.
Therefore, the relevant input for this calculation is the recombination
matrix element between the baryon state, which has a pair of quasi-particle
excitations of momentum-spin labels $\left(\mathbf{p},\mathrm{up};\mathbf{p}^{\prime},\mathrm{down}\right)$,
and the same state but with these excitations restored to the condensate.
To the leading (zero) order in $k\ll p_{F}$, this matrix element
is usually estimated as\begin{equation}
\left\vert \mathcal{M}_{B}\right\vert ^{2}=\frac{\Delta^{2}}{\epsilon_{p}^{2}}\label{ME}\end{equation}
 yielding the following neutrino energy losses at temperature $T<T_{c}$.\begin{equation}
Q_{\mathrm{FRS}}\left(^{1}S_{0}\right)=\frac{4G_{F}^{2}{p}_{F}M^{\ast}C_{V}^{2}}{15\pi^{5}}\mathcal{N}_{\nu}T^{7}y^{2}\int_{0}^{\infty}\;\frac{z^{4}dx}{\left(e^{z}+1\right)^{2}},\label{SF}\end{equation}
 where $M^{\ast}$ is the effective nucleon mass, $y=\Delta/T$, $z=\sqrt{x^{2}+y^{2}}$,
and $\mathcal{N}_{\nu}=3$ is the number of neutrino flavors. $T_{c}$
is the critical temperature for baryon pairing.

The naive estimate (\ref{ME}) is inconsistent with the hypothesis
of conservation of the vector current in weak interactions. Indeed,
a longitudinal vector current of quasi-particles consisting only on
a temporal component can not satisfy the continuity equation. 

The question of conservation of the longitudinal vector current caused
by recombination of quasi-particles has been discussed by many people
\cite{many} in connection with the gauge invariance of the Bardeen-Cooper-Schrieffer
theory of superconductivity. It was realized that the current conservation
could be restored if the interaction among quasi-particles is incorporated
in the coupling vertex to the same degree of approximation as the
self-energy effect is included in the quasi-particle. It has been
also pointed out that there is significant difference between the
transverse and longitudinal current operators in their matrix elements.
Namely, there exist collective excited states of quasi-particle pairs
\cite{Bogoliubov}, \cite{Nambu} which can be excited only by the
longitudinal current. As a result, the spatial part of the longitudinal
current does not vanish, and cancels the temporal part.

In the present paper we recalculate the neutrino energy losses with
allowance for conservation of the weak vector current. The paper is
organized as follows. In section 2 we shortly discuss the Nambu-Gorkov
formalism as a convenient description of the particle-hole excitations
in the system with pairing. The wave functions of the quasi-particle
excitations are obtained in section 3. In section 4 we derive an effective
vertex conserving the vector weak current for neutral and charged
baryons. In section 5 we use the quasi-particle states for the calculation
of the matrix element of the effective vector weak current, and the
neutrino energy losses in the vector channel. The neutrino energy
losses in the axial channel are calculated in section 6. A discussion
of the obtained results and our main conclusions are presented in
section 7.

\section{Formalism}

In the Nambu-Gorkov formalism, the quasi-particle fields are represented
by two-component objects\begin{equation}
\Psi_{p}=\left(\begin{array}{c}
\psi_{\mathbf{1}}\left(p\right)\\
\psi_{2}^{\dagger}\left(-p\right)\end{array}\right)\label{tcf}\end{equation}
 $\psi_{\mathbf{1}}\left(p\right)$ is the the quasi-particle component
of the excitation with momentum $\mathbf{p}$ and spin $\sigma$,
and $\psi_{2}^{\dagger}\left(-p\right)$ is the hole component of
the same excitation, which can be interpreted as the absence of a
particle with momentum $-\mathbf{p}$ and spin $-\sigma$. The two-component
fields (\ref{tcf}) obey the standard fermion commutation relations
\[
\left\{ \Psi_{p,\sigma},\Psi_{p^{\prime},\sigma^{\prime}}\right\} =\delta_{\sigma,\sigma^{\prime}}\delta_{p,p^{\prime}}.\]
 With the aid of the $2\times2$ Pauli matrices\begin{equation}
\hat{\tau}_{1}=\left(\begin{array}{cc}
0 & 1\\
1 & 0\end{array}\right),\ \hat{\tau}_{2}=\left(\begin{array}{cc}
0 & -i\\
i & 0\end{array}\right),\ \hat{\tau}_{3}=\left(\begin{array}{cc}
1 & 0\\
0 & -1\end{array}\right)\label{tau}\end{equation}
 operating in the particle-hole space, the Hamiltonian of the system
of quasi-particles can be recast as \cite{Nambu}\[
H=H_{0}+H_{1},\]
 where \begin{equation}
H_{0}=\sum_{p}\Psi_{p}^{\dagger}\left(\xi_{\mathbf{p}}\hat{\tau}_{3}+\hat{\Lambda}\left(\mathbf{\check{p}}\right)\right)\Psi_{p}\label{Ham}\end{equation}
 is the BCS reduced Hamiltonian, where the nonrelativistic energy
is measured relatively to the Fermi level\[
\xi_{\mathbf{p}}\equiv\frac{p^{2}}{2M^{\ast}}-\mu,\]
 $M^{\ast}$ is the effective mass of the quasi-particle and $\mu$
is the Fermi energy. The quasi-particle self-energy, of the form\begin{equation}
\hat{\Lambda}\left(\mathbf{\check{p}}\right)\equiv\left(\begin{array}{cc}
0 & \hat{D}\left(\mathbf{\check{p}}\right)\\
\hat{D}^{\dagger}\left(\mathbf{\check{p}}\right) & 0\end{array}\right),\label{Lam}\end{equation}
 is a $2\times2$ matrix in the Nambu-Gorkov space, and a $2\times2$
matrix in the spin space, which depends on the orientation of the
quasi-particle momentum $\mathbf{\check{p}}=\mathbf{p}/\left\vert \mathbf{p}\right\vert $
(see e.g. \cite{Tamagaki}).

The residual interaction among quasi-particles is given by the following
Hamiltonian\[
H_{1}=\frac{1}{2}\sum_{p,p^{\prime}}V_{pp^{\prime}}\left(q\right)\left(\Psi_{p+q}^{\dagger}\hat{\tau}_{3}\Psi_{p}\right)\left(\Psi_{p^{\prime}-q}^{\dagger}\hat{\tau}_{3}\Psi_{p\prime}\right).\]

As follows from the Hamiltonian (\ref{Ham}), the inverse of the quasi-particle
propagator has the simple form:\begin{equation}
i\text{\c{G}}^{-1}\left(p\right)=p_{0}-\xi_{\mathbf{p}}\hat{\tau}_{3}-\hat{\Lambda}\left(\mathbf{\check{p}}\right)\label{iG}\end{equation}

\section{Quasi-particle states}

Near the Fermi-surface, the imaginary part of the quasi-particle energy
is small. Therefore, to the extent that the single-particle picture
makes some physical sense, we will describe the quasi-particles with
the aid of wave-functions. The states of quasi-particles obey the
equation\begin{equation}
\text{\c{G}}^{-1}\Psi_{\mathbf{p}}=0.\label{e}\end{equation}
 Writing $\Psi_{\mathbf{p}}$ in the form\begin{equation}
\Psi_{\mathbf{p}}\equiv\frac{1}{\sqrt{N}}\left(\begin{array}{c}
\alpha\\
\beta\end{array}\right)\label{ab}\end{equation}
 with $\alpha$ and $\beta$ being two-component spinors, we arrive
to the following set of matrix equations\begin{eqnarray*}
\left(p_{0}-\xi_{p}\right)\alpha-\hat{D}\beta & = & 0\\
\hat{D}^{\dagger}\alpha-\left(p_{0}+\xi_{p}\right)\beta & = & 0\end{eqnarray*}
 From the second equation we have \begin{equation}
\alpha=\left(p_{0}+\xi_{p}\right)\left(\hat{D}^{\dagger}\right)^{-1}\beta\label{alpha}\end{equation}
 By substituting this into the first equation and multiplying $\hat{D}^{\dagger}$
from the left, we obtain\begin{equation}
\left(p_{0}^{2}-\xi_{p}^{2}-\hat{D}^{\dagger}\hat{D}\right)\beta=0\label{beta}\end{equation}
 This equation has nontrivial solutions only if \begin{equation}
\det\left\vert p_{0}^{2}-\xi_{p}^{2}-\hat{D}^{\dagger}\hat{D}\right\vert =0.\label{det}\end{equation}
 In the interesting cases of singlet- and triplet-state pairing we
are going to consider, the gap matrix $\hat{D}$ is proportional to
the unitary matrix $\hat{\Upsilon}\left(\mathbf{\check{p}}\right)$\begin{equation}
\hat{D}\left(\mathbf{\check{p}}\right)=\Delta_{\mathbf{\check{p}}}\hat{\Upsilon}\left(\mathbf{\check{p}}\right)\label{DG}\end{equation}
 with $\Delta_{\mathbf{\check{p}}}$

a scalar function \cite{Tamagaki}, so that \begin{equation}
\hat{\Upsilon}^{\dagger}\hat{\Upsilon}=1\label{Un}\end{equation}
 Eq. (\ref{det}) then gives us the eigenvalues\begin{equation}
p_{0}=\pm\epsilon_{\mathbf{p}},\ \ \ \ \epsilon_{\mathbf{p}}\equiv\sqrt{\xi_{p}^{2}+\Delta_{\mathbf{\check{p}}}^{2}}\label{p0}\end{equation}
 with \begin{equation}
\Delta_{\mathbf{\check{p}}}^{2}=\frac{1}{2}Tr\hat{D}^{\dagger}\hat{D}\label{Dlam}\end{equation}

Using the fact that Eq. (\ref{beta}) is diagonal, we may choose two
independent solutions for $\beta$ as the ordinary spinors: \begin{equation}
\beta_{1}^{\pm}=\chi_{\uparrow}\equiv\left(\begin{array}{c}
1\\
0\end{array}\right),\ \ \ \beta_{2}^{\pm}=\chi_{\downarrow}\equiv\left(\begin{array}{c}
0\\
1\end{array}\right)\label{bpm}\end{equation}
 Here, the upper sign corresponds to the positive- or negative-frequency
solution, in accordance with Eq. (\ref{p0}). The $\alpha$-component
of these solutions is to be found from Eq. (\ref{alpha}).

From Eq. (\ref{DG}), we obtain\[
\left(\hat{D}^{\dagger}\right)^{-1}=\Delta_{\mathbf{\check{p}}}^{-1}\hat{\Upsilon}\left(\mathbf{\check{p}}\right)\]
 Substitution this into Eq. (\ref{alpha}) gives \begin{equation}
\alpha_{i}^{\pm}=\Delta_{\mathbf{\check{p}}}^{-1}\left(\pm\epsilon_{\mathbf{p}}+\xi_{p}\right)\hat{\Upsilon}\beta_{i}\label{apm}\end{equation}

Making use of Eqs. (\ref{bpm}), (\ref{apm}) we can find the normalized
wave functions. The positive-frequency states can be written as\begin{equation}
\Psi_{\mathbf{p},\sigma}=\frac{1}{\sqrt{N}}\left(\begin{array}{c}
\Delta_{\mathbf{\check{p}}}^{-1}\left(\epsilon_{\mathbf{p}}+\xi_{p}\right)\hat{\Upsilon}\chi_{\sigma}\\
\chi_{\sigma}\end{array}\right)e^{i\mathbf{pr}-i\epsilon_{\mathbf{p}}t}.\label{vf}\end{equation}
 The factor $1/\sqrt{N}$, in Eq. (\ref{ab}) is to be found from
the normalization condition $\Psi_{\mathbf{p}}^{\dagger}\Psi_{\mathbf{p}}=1$.
A direct evaluation making use of Eq. (\ref{Un}) yields\[
\frac{1}{\sqrt{N}}=\sqrt{\frac{\epsilon_{p}-\xi_{p}}{2\epsilon_{p}}}.\]
 Incorporating this factor into Eq. (\ref{vf}) we obtain the positive-frequency
eigenstates\begin{equation}
\Psi_{\mathbf{p},\sigma}=\left(\begin{array}{c}
u_{\mathbf{p}}\hat{\Upsilon}\chi_{\sigma}\\
v_{\mathbf{p}}\chi_{\sigma}\end{array}\right)e^{i\mathbf{pr}-i\epsilon_{\mathbf{p}}t}\label{pvf}\end{equation}
 with\begin{equation}
u_{\mathbf{p}}=\sqrt{\frac{\epsilon_{p}+\xi_{p}}{2\epsilon_{p}}},\ \ \ \
v_{\mathbf{p}}=\sqrt{\frac{\epsilon_{p}-\xi_{p}}{2\epsilon_{p}}}.\label{up}\end{equation}
 The negative-frequency wave-functions can be obtained in the same
way. The state with momentum $\mathbf{p}$ and spin-label $\sigma$
has the form \begin{equation}
\Psi_{-\mathbf{p},-\sigma}=\left(\begin{array}{c}
-v_{\mathbf{p}}\hat{\Upsilon}\chi_{-\sigma}\\
u_{\mathbf{p}}\chi_{-\sigma}\end{array}\right)e^{-i\mathbf{pr}+i\epsilon_{\mathbf{p}}t}\label{nvf}\end{equation}
 This solution is connected to the hole state by the particle-antiparticle
conjugation\[
C:\ \Psi^{C}=C\Psi^{\dagger}=\hat{\tau}_{2}\Psi^{\dagger}.\]
 which changes quasi-particles of energy-momentum $\left(p_{0},\mathbf{p}\right)$
into holes of energy-momentum $\left(-p_{0},-\mathbf{p}\right)$,
or interchanges up-spin and down-spin particles.

It is necessary to note that, in general, $\Psi_{\mathbf{p},\sigma}$
is not an eigenstate of the spin projection, because the upper component
$\propto\hat{\Upsilon}\chi_{\sigma}$ is a linear combination of different
spin states. The label $\sigma$ indicates only the spin function
of the lower component, and serves for identification of the quasi-particle
states.

\section{Effective vertex for quasi-particles}

\subsection{Neutral baryons}

The components of the bare vertex \begin{equation}
\gamma^{\mu}=\left\{ \begin{array}{cc}
\hat{\tau}_{3} & \mathrm{if\ }\mu=0,\ \ \ \ \ \ \ \ \ \ \ \ \\
\frac{1}{M^{\ast}}\mathbf{p} & \mathrm{if\ }\mu=i=1,2,3\end{array}\right..\label{gam}\end{equation}
 are $2\times2$ matrices in the Nambu-Gorkov space. As already mentioned,
the longitudinal current corresponding to the bare vertex does not
satisfy the continuity equation. To restore the current conservation,
one must consider the modification of the vertex $\gamma^{\mu}$ to
the same order as the modification of the propagator is done. The
relation between the modified vertex $\Gamma^{\mu}$ and the quasi-particle
propagator (\ref{iG}) is given by the Ward identity \cite{Schr}
\begin{equation}
K_{\mu}\Gamma^{\mu}\left(p^{\prime},p\right)=\hat{\tau}_{3}\text{\c{G}}^{-1}\left(p\right)-\text{\c{G}}^{-1}\left(p^{\prime}\right)\hat{\tau}_{3},\label{Ward}\end{equation}
 where $K=\left(\omega,\mathbf{k}\right)$ is the transferred momentum.
The plane wave solutions \[
u_{\mathbf{p},\alpha}\exp\left(i\mathbf{pr}-i\epsilon_{\mathbf{p}}t\right),\ \ \ \ \ u_{\mathbf{p}^{\prime},\alpha^{\prime}}^{\ast}\exp\left(-i\mathbf{pr}+i\epsilon_{\mathbf{p}}t\right)\]
 for $\Psi$ and $\Psi^{+}$ obey the equations \c{G}$^{-1}\left(p\right)u_{\mathbf{p},\alpha}=0$,
and $u_{\mathbf{p}^{\prime},\alpha^{\prime}}^{\ast}$\c{G}$^{-1}\left(p^{\prime}\right)=0$.
Therefore the Ward identity implies conservation of the vector current
on the energy shell of the quasi-particles.

Following the prescriptions of quantum electrodynamics, an approximation
which satisfies the Ward identity (and hence the continuity equation)
is the sum of ladder diagrams satisfying the Dyson equation shown
in Fig. 1.

\begin{figure}
\includegraphics{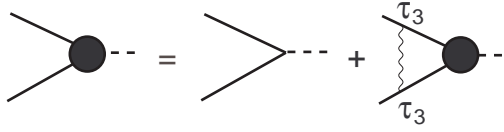}

\caption{Dyson equation, in the ladder approximation, for the effective vertex.
Solid lines stand for dressed particles, and the dashed line represents
the $Z$-boson field. The wavy line accounts for the residual interaction
among quasi-particles.}
\end{figure}

In these diagrams, all solid lines are \char`\"{}dressed\char`\"{}
particles, and the dashed line is the $Z$-boson field. The \char`\"{}dressed\char`\"{}
particles interact with the same primary interaction $V_{pp^{\prime}}$
which produces the self-energy of the quasi-particle. Thus, for $K=p-p^{\prime}$,
the corrected vertex satisfies the integral equation\begin{eqnarray}
\Gamma^{\mu}\left(p-K,p\right) & = & \hat{\gamma}^{\mu}\left(p-K,p\right)\label{Gam}\\
 &  & +i\int\frac{d^{4}p^{\prime}}{\left(2\pi\right)^{4}}\ \hat{\tau}_{3}\text{\c{G}}\left(p^{\prime}-K\right)\Gamma^{\mu}\left(p^{\prime}-K,p^{\prime}\right)\text{\c{G}}\left(p^{\prime}\right)\hat{\tau}_{3}V_{pp^{\prime}}\notag\end{eqnarray}

In the limit $K=\left(\omega,\mathbf{0}\right)$, the Ward identity
gives \[
\Gamma^{0}\left(p-K,p\right)=\hat{\tau}_{3}-\frac{2}{\omega}\hat{\tau}_{3}\hat{\Lambda}\left(\mathbf{\check{p}}\right)\]
 with\begin{equation}
\hat{\tau}_{3}\hat{\Lambda}\left(\mathbf{\check{p}}\right)=\left(\begin{array}{cc}
0 & \hat{D}\left(\mathbf{\check{p}}\right)\\
-\hat{D}^{\dagger}\left(\mathbf{\check{p}}\right) & 0\end{array}\right)\label{A}\end{equation}
 The poles of the vertex function correspond to collective eigen-modes
of the system. Therefore, the pole which appears at $\omega\rightarrow0$,
$k=0$ implies the existence of a collective mode, which plays an
important role in the conservation of the vector current. The corresponding
nonperturbative solution to Eq. (\ref{Gam}) has been found by Nambu
\cite{Nambu} (see also \cite{Littlewood}). In our notation, it reads%
\footnote{To obtain the weak vector current, this vertex should be multiplied
by the weak coupling constant $C_{V}$.%
} \begin{equation}
\Gamma_{0}\left(p-K,p\right)=\hat{\tau}_{3}-2\hat{\tau}_{3}\hat{\Lambda}\left(\mathbf{\check{p}}\right)\frac{\omega}{\omega^{2}-a^{2}k^{2}}\label{G0}\end{equation}
 \begin{equation}
\mathbf{\Gamma}=\frac{\mathbf{p}}{M}-2\hat{\tau}_{3}\hat{\Lambda}\left(\mathbf{\check{p}}\right)\frac{a^{2}\mathbf{k}}{\omega^{2}-a^{2}k^{2}},\label{GV}\end{equation}
 The poles in this vertex correspond to the collective motion of the
condensate, with the dispersion relation $\omega=ak$, where $a^{2}=V_{F}^{2}/3$.

The effective vertex satisfies the Ward identity (\ref{Ward}), and
thus the continuity equation on the energy shell\begin{equation}
\omega\Gamma_{0}-\mathbf{k\Gamma}\simeq0\label{cvc}\end{equation}
 Indeed, by making use of $\omega=\epsilon_{\mathbf{p}}+\epsilon_{\mathbf{p}^{\prime}}$
and $\mathbf{k}=\mathbf{p}+\mathbf{p}^{\prime}$ with $\ k\ll p\simeq p_{F}$,
the left-hand side of this equation can be reduced to the right-hand
side of the Ward identity (\ref{Ward}).

\subsection{Charged baryons}

Consider now the case when quasi-particles carry an electric charge.
Including the long-range Coulomb interaction $V_{C}\left(k\right)=e^{2}/k^{2}$
implies that the vertex part is multiplied by a string of closed loops,
which represents the polarization of the surrounding medium. In this
case, the new vertex $\tilde{\Gamma}^{\mu}$ can be found as the solution
of the Dyson equation, according to the diagram of Fig. 2. or, analytically
\begin{eqnarray}
\tilde{\Gamma}^{\mu}\left(p-K,p\right) & = & \Gamma^{\mu}\left(p-K,p\right)-\Gamma_{0}\left(p-K,p\right)V_{C}\left(k\right)\notag\\
 &  & \times i\int\frac{d^{4}p^{\prime}}{\left(2\pi\right)^{4}}\ Tr\left[\hat{\tau}_{3}\text{\c{G}}\left(p^{\prime}-K\right)\tilde{\Gamma}^{\mu}\left(p^{\prime}-K,p^{\prime}\right)\text{\c{G}}\left(p^{\prime}\right)\right]\label{gc}\end{eqnarray}
\begin{figure}
\includegraphics{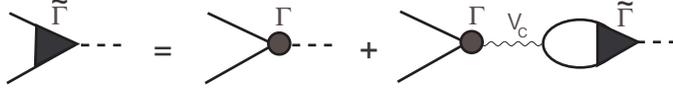}

\caption{Dyson equation for the vertex correction for charged quasi-particles
(compare with Fig. 1). The shaded areas represent the modified effective
vertex, and the wavy line stands for the Coulomb interaction. }
\end{figure}

The integral\begin{equation}
\tilde{\Pi}^{0\mu}\left(K\right)\equiv i\int\frac{d^{4}p^{\prime}}{\left(2\pi\right)^{4}}\ Tr\left[\hat{\tau}_{3}\text{\c{G}}\left(p^{\prime}-K\right)\tilde{\Gamma}^{\mu}\left(p^{\prime}-K,p^{\prime}\right)\text{\c{G}}\left(p^{\prime}\right)\right]\label{Pt}\end{equation}
 in the r.h.s. of Eq. (\ref{gc}) represents the $\tilde{\Pi}^{0\mu}$
component of the polarization tensor shown by the loop in Fig. 2.
Introducing also the notation\[
\Pi^{0\mu}\left(K\right)\equiv i\int\frac{d^{4}p^{\prime}}{\left(2\pi\right)^{4}}\ Tr\left[\hat{\tau}_{3}\text{\c{G}}\left(p^{\prime}-K\right)\Gamma^{\mu}\left(p^{\prime}-K,p^{\prime}\right)\text{\c{G}}\left(p^{\prime}\right)\right]\]
 Eq. (\ref{gc}) can be recast as follows\[
\tilde{\Pi}^{0\mu}\left(K\right)=\Pi^{0\mu}\left(K\right)-V_{C}\left(k\right)\Pi^{00}\left(K\right)\tilde{\Pi}^{0\mu}\left(K\right).\]
 We then obtain\[
\tilde{\Pi}^{0\mu}\left(K\right)=\frac{\Pi^{0\mu}\left(K\right)}{1+V_{C}\left(k\right)\Pi^{00}\left(K\right)}.\]
 Combining this with Eqs. (\ref{gc}), (\ref{Pt}) yields\[
\tilde{\Gamma}^{\mu}\left(p-K,p\right)=\Gamma^{\mu}\left(p-K,p\right)\left(1-\frac{V_{C}\left(k\right)\Pi^{0\mu}\left(K\right)}{1+V_{C}\left(k\right)\Pi^{00}\left(K\right)}\right).\]
 In particular, for $\Gamma^{0}$ we arrive to%
\footnote{The solution to the equation $1+V_{C}\left(k\right)\Pi^{00}\left(K\right)=0$
determines the new dispersion law $\omega=\omega\left(k\right)$ for
the collective excitations, which represents plasma waves \cite{Nambu}.%
}\begin{equation}
\tilde{\Gamma}^{0}\left(p-K,p\right)=\frac{\Gamma^{0}\left(p-K,p\right)}{1+V_{C}\left(k\right)\Pi^{00}\left(K\right)}.\label{G0em}\end{equation}

The polarization function $\Pi^{00}\left(K\right)$ can be readily
calculated with the help of $\Gamma^{0}$ given by Eq. (\ref{G0}).
By neglecting the small dependence of the energy gap on the transferred
momentum $\mathbf{k}$, we have\begin{eqnarray*}
\Pi^{00}\left(K\right) & = & i\int\frac{d^{4}p^{\prime}}{\left(2\pi\right)^{4}}\ Tr\left[\hat{\tau}_{3}\text{\c{G}}\left(p^{\prime}-K\right)\hat{\tau}_{3}\text{\c{G}}\left(p^{\prime}\right)\right]\\
 &  & -\frac{2\omega}{\omega^{2}-a^{2}k^{2}}i\int\frac{d^{4}p^{\prime}}{\left(2\pi\right)^{4}}\ Tr\left[\hat{\tau}_{3}\text{\c{G}}\left(p^{\prime}-K\right)\hat{\tau}_{3}\hat{\Lambda}\left(p^{\prime}\right)\text{\c{G}}\left(p^{\prime}\right)\right].\end{eqnarray*}
 Here the quasi-particle propagator follows from Eq. (\ref{iG}):
\begin{equation}
\text{\c{G}}\left(p\right)=\frac{i}{p_{0}^{2}-\epsilon_{p}^{2}}\left(p_{0}+\xi_{p}\hat{\tau}_{3}+\hat{\Lambda}\left(\mathbf{\check{p}}\right)\right)\label{Gp}\end{equation}
 Performing the trace over particle-hole and spin indices, and integration
over $p_{0}$, results in the following expression\begin{eqnarray*}
\Pi^{00}\left(\omega,\mathbf{k}\right) & = & \int\frac{2d^{3}p}{\left(2\pi\right)^{3}}\ \frac{\omega\epsilon_{\mathbf{p}-\mathbf{k}}-\Delta_{\mathbf{\check{p}}}^{2}-\xi_{\mathbf{p}}\xi_{\mathbf{p-k}}-\epsilon_{\mathbf{p-k}}^{2}-a^{2}k^{2}\left(\omega^{2}-a^{2}k^{2}\right)^{-1}2\Delta_{\mathbf{\check{p}}}^{2}}{\epsilon_{\mathbf{p}-\mathbf{k}}\left(\omega-\epsilon_{\mathbf{p}}-\epsilon_{\mathbf{p-k}}\right)\left(\omega+\epsilon_{\mathbf{p}}-\epsilon_{\mathbf{p-k}}\right)}\\
 &  & +\int\frac{2d^{3}p}{\left(2\pi\right)^{3}}\ \frac{-\omega\epsilon_{\mathbf{p}}-\Delta_{\mathbf{\check{p}}}^{2}-\epsilon_{\mathbf{p}}^{2}-\xi_{\mathbf{p}}\xi_{\mathbf{p-k}}-a^{2}k^{2}\left(\omega^{2}-a^{2}k^{2}\right)^{-1}2\Delta_{\mathbf{\check{p}}}^{2}}{\epsilon_{\mathbf{p}}\left(\omega+\epsilon_{\mathbf{p}}+\epsilon_{\mathbf{p-k}}\right)\left(\omega-\epsilon_{\mathbf{p-k}}+\epsilon_{\mathbf{p}}\right)}\end{eqnarray*}
 We are interested in the regime defined by \[
k<\omega,\ \ \ \omega>2\epsilon_{\mathbf{p}}\gg\epsilon_{\mathbf{p}}-\epsilon_{\mathbf{p-k}}\simeq kV_{F}\]
 In this case we obtain, after some simplifications%
\footnote{An analogous expression has been obtained in \cite{Nambu} for the
case of singlet pairing of electrons.%
}

\begin{eqnarray*}
\Pi^{00}\left(\omega,\mathbf{k}\right) & = & \frac{1}{8\pi^{3}}\left[\frac{a^{2}k^{2}}{\omega^{2}-a^{2}k^{2}}\int d^{3}p\ \frac{\Delta_{\mathbf{\check{p}}}^{2}}{\epsilon_{p}\left(\epsilon_{p}^{2}-a^{2}k^{2}/4\right)}\right.\\
 &  & \left.+k^{2}\int d^{3}p\ \frac{\Delta_{\mathbf{\check{p}}}^{2}}{\epsilon_{p}\left(\omega^{2}-4\epsilon_{p}^{2}\right)}\left(\frac{p^{2}}{3M^{2}\epsilon_{p}^{2}}-\frac{a^{2}}{\epsilon_{p}^{2}-a^{2}k^{2}/4}\right)\right].\end{eqnarray*}
 with $a^{2}=V_{F}^{2}/3$.

Inserting this into Eq. (\ref{G0em}) gives\begin{equation}
\tilde{\Gamma}^{0}\left(p-K,p\right)=\frac{\Gamma^{0}\left(p-K,p\right)}{1+\chi\left(K\right)}\label{G0em1}\end{equation}
 with\begin{eqnarray}
\chi\left(K\right) & = & \frac{e^{2}}{8\pi^{3}}\left[\frac{a^{2}}{\omega^{2}-a^{2}k^{2}}\int d^{3}p\ \allowbreak\frac{\Delta_{\mathbf{\check{p}}}^{2}}{\epsilon_{p}\left(\epsilon_{p}^{2}-a^{2}k^{2}/4\right)}\right.\notag\\
 &  & \left.+\int d^{3}p\ \frac{\Delta_{\mathbf{\check{p}}}^{2}}{\epsilon_{p}\left(\omega^{2}-4\epsilon_{p}^{2}\right)}\allowbreak\left(\frac{p^{2}}{3M^{2}\epsilon_{p}^{2}}-\frac{a^{2}}{\epsilon_{p}^{2}-a^{2}k^{2}/4}\right)\right].\label{hi}\end{eqnarray}

\section{Energy losses in the vector channel}

\subsection{Neutral baryons}

Having at hand the effective vertex and the wave functions of quasi-particles
and holes, we can evaluate the matrix element of the vector weak current.
In the particle-hole picture, the creation and recombination of two
quasi-particles is described by the off-diagonal matrix elements of
the Hamiltonian, which corresponds to quasi-particle transitions into
a hole (and a correlated pair). Thus, we calculate the matrix element
of the current between the initial (positive-frequency) state of a
quasi-particle with momentum $\mathbf{p}$ and the final (negative-frequency)
state with the same momentum $\mathbf{p}$. \[
\mathcal{M}_{\mu}=\left\langle \Psi_{-\mathbf{p,}-\sigma}^{\dagger}\left\vert \Gamma_{\mu}\right\vert \Psi_{\mathbf{p,\sigma}}\right\rangle \]

Let us consider separately the contributions from the bare vertex,
given by the first terms in Eq. (\ref{G0}), and the collective part,
given by the second term, so that $\mathcal{M}_{\mu}=\mathcal{M}_{\mu}^{\mathrm{bare}}+\mathcal{M}_{\mu}^{\mathrm{coll}}$.

Making use of the wave functions described by Eqs. (\ref{pvf}), (\ref{nvf})
and using the identity (\ref{Un}), for $\mu=0$, we find \begin{equation}
\mathcal{M}_{0}^{\mathrm{bare}}=-\left(u_{p}v_{p^{\prime}}+v_{p}u_{p^{\prime}}\right)\simeq-\frac{\Delta_{\mathbf{\check{p}}}}{\epsilon_{p}},\ \ \ \ k\ll p\simeq p_{F}\label{first}\end{equation}
 \[
\mathcal{M}_{0}^{\mathrm{coll}}=2\Delta_{\mathbf{\check{p}}}\frac{\omega}{\omega^{2}-a^{2}k^{2}}\left(u_{p}u_{p^{\prime}}+v_{p}v_{p^{\prime}}\right)\simeq2\Delta_{\mathbf{\check{p}}}\frac{\omega}{\omega^{2}-a^{2}k^{2}}.\]
 with $\epsilon_{p}+\epsilon_{p^{\prime}}=\omega$ and $\mathbf{p+p}^{\prime}=\mathbf{k}$.

The velocity of the collective mode $a^{2}=V_{F}^{2}/3$ is small
in the nonrelativistic system. Therefore, we expand the collective
contribution in this parameter to obtain\begin{equation}
\mathcal{M}_{0}^{\mathrm{coll}}\simeq\frac{\Delta_{\mathbf{\check{p}}}}{\epsilon_{p}}\left(1+\frac{1}{3}V_{F}^{2}\frac{k^{2}}{\omega^{2}}\right)\label{second}\end{equation}

The contribution of the bare vertex $\mathcal{M}_{0}^{\mathrm{bare}}$
reproduces the matrix element (\ref{ME}) derived by Flowers et al.
\cite{FRS76} and Yakovlev et al. \cite{YKL98}. However, the collective
correction modifies this crucially. In the sum of the two contributions,
the leading terms mutually cancel, yielding the matrix element\[
\mathcal{M}_{0}=\mathcal{M}_{0}^{\mathrm{bare}}+\mathcal{M}_{0}^{\mathrm{coll}}\simeq\frac{1}{3}V_{F}^{2}\frac{k^{2}}{\omega^{2}}\frac{\Delta_{\mathbf{\check{p}}}}{\epsilon_{p}}\]
 which is at least $\sim V_{F}^{2}$ times smaller than the bare result.

The above cancellation of the contribution of \char`\"{}bare\char`\"{}
vertex is not accidental. This part of the matrix element, which remains
finite at zero transferred momentum, cannot be restored with taking
into account of more complicated corrections to the vertex part. It
is well-known that, for any process involving interactions, the perturbation
diagrams can be grouped into gauge invariant subsets, such that the
invariance is maintained by each subset taken as a whole. If any new
subset of the diagrams is added to the vertex part, the same should
also be incorporated into the quasi-particle self-energy. If this
is carefully done, for $k\rightarrow0$, one must obtain the exact
cancellation of the contribution of the \char`\"{}bare\char`\"{} vertex.
Indeed, $\Psi^{+}\left(\mathbf{r},t\right)\Psi\left(\mathbf{r},t\right)$
represents the operator of baryon charge density. If the ground state
$\left\vert 0\right\rangle $ is an eigenstate of the baryon charge
operator for the system with $N$ particles, and $|n>$ is a state
for the system with $N+2$ particles, the matrix element\[
\int\left\langle 0\right\vert \Psi_{B}^{+}\left(\mathbf{r},t\right)\Psi_{B}\left(\mathbf{r},t\right)\left\vert n\right\rangle e^{i\mathbf{kr}-i\omega t}d^{3}rdt\]
 should vanish for $k\rightarrow0$, because the above expression
is an off-diagonal matrix element of the total charge operator $Q$.
This implies that, for $k\ll p_{F}$ , the matrix element should be
proportional to some power of $k$.

The spatial component of the longitudinal (with respect to $\mathbf{k}$)
component of the matrix element can be obtained from Eq. (\ref{cvc}).
Since $\mathbf{\check{k}\Gamma=}\left(\omega/k\right)\Gamma_{0}$
we have\[
\mathcal{M}_{\parallel}\simeq\frac{1}{3}V_{F}^{2}\frac{k}{\omega}\frac{\Delta_{\mathbf{\check{p}}}}{\epsilon_{p}}.\]
 In the above, $\ \mathbf{\check{k}=k}/k$ is a unit vector directed
along the transferred momentum.

Since the collective interaction modifies only the longitudinal part
of the vertex, the transverse part of the matrix element can be evaluated
directly from the bare vertex (\ref{gam}). This yields\[
\mathcal{M}_{\perp}\simeq\left(v_{p}u_{p^{\prime}}-u_{p}v_{p^{\prime}}\right)\frac{\mathbf{p}_{\perp}}{M^{\ast}}\simeq-\frac{1}{2}V_{F}^{2}\frac{k\Delta_{\mathbf{\check{p}}}}{\epsilon_{\mathbf{p}}^{2}}\left(\mathbf{\check{k}\check{p}}\right)\mathbf{\check{p}}_{\perp}.\]

The rate of the process is proportional to the square of the matrix
element. This means that the vector current contribution to the neutrino
energy losses is $V_{F}^{4}$ times smaller than estimated before.
The corresponding neutrino emissivity in the vector channel can be
evaluated with the aid of Fermi's golden rule: \begin{eqnarray*}
Q_{V} & = & \left(\frac{G_{F}}{2\sqrt{2}}\right)^{2}\frac{C_{V}^{2}}{\left(2\pi\right)^{8}}\mathcal{N}_{\nu}\int d^{3}pd^{3}p^{\prime}f\left(\epsilon_{\mathbf{p}}\right)f\left(\epsilon_{\mathbf{p}^{\prime}}\right)\\
 &  & \times\int\frac{d^{3}k_{1}}{2\omega_{1}}\frac{d^{3}k_{2}}{2\omega_{2}}\omega\mathrm{Tr}\left(l_{\mu}l_{\nu}^{\ast}\right)\mathcal{M}^{\mu}\mathcal{M}^{\nu}\delta\left(\mathbf{p+p}^{\prime}-\mathbf{k}\right)\delta\left(\epsilon_{\mathbf{p}}+\epsilon_{\mathbf{p}^{\prime}}-\omega\right).\end{eqnarray*}
 One can simplify this equation by inserting $\int d^{4}K\ \delta^{\left(4\right)}\left(K-k_{1}-k_{2}\right)=1$.
Then, the phase-space integrals for neutrinos are readily done with
the aid of Lenard's formula \begin{eqnarray*}
 &  & \int\frac{d^{3}k_{1}}{2\omega_{1}}\frac{d^{3}k_{2}}{2\omega_{2}}\delta^{\left(4\right)}\left(K-k_{1}-k_{2}\right)\mathrm{Tr}\left(j^{\mu}j^{\nu\ast}\right)\\
 & = & \frac{4\pi}{3}\left(K_{\mu}K_{\nu}-K^{2}g_{\mu\nu}\right)\Theta\left(K^{2}\right)\Theta\left(\omega\right),\end{eqnarray*}
 where $\Theta(x)$ is the Heaviside step function.

For $k\ll p_{F}$ we obtain\begin{eqnarray*}
Q_{V} & = & \frac{4\pi}{3}\left(\frac{G_{F}}{2\sqrt{2}}\right)^{2}\frac{C_{V}^{2}}{\left(2\pi\right)^{8}}\mathcal{N}_{\nu}\int d^{3}pf^{2}\left(\epsilon_{\mathbf{p}}\right)\int_{0}^{\infty}d\omega\omega\int_{0}^{\omega}dkk^{2}d\Omega_{k}\\
 &  & \times\left(\left(K_{\mu}\mathcal{M}^{\mu}\right)^{2}-K^{2}\mathcal{M}^{\mu}\mathcal{M}_{\mu}\right)\delta\left(2\epsilon_{\mathbf{p}}-\omega\right).\end{eqnarray*}
 The next integrations are trivial. We get\[
Q_{V}=\frac{592}{42\,525\pi^{5}}V_{F}^{4}G_{F}^{2}C_{V}^{2}p_{F}M^{\ast}T^{7}y^{2}\int_{0}^{\infty}\;\frac{z^{4}dx}{\left(e^{z}+1\right)^{2}}\]
 This is to be compared with Eq. (\ref{SF}). We see that \[
\frac{Q_{V}}{Q_{\mathrm{FRS}}\left(_{1}S^{0}\right)}=\frac{148}{2835}V_{F}^{4}\]
 i.e. the neutrino radiation via the vector weak currents in the nonrelativistic
system $\left(V_{F}\ll1\right)$ is suppressed by several orders of
magnitude with respect to the predictions of Flowers et al. \cite{FRS76}.

\subsection{Charged baryons}

If the paired baryons carry an electric charge, the effective vector
vertex is given by Eqs. (\ref{G0em1}), (\ref{G0}) and (\ref{hi}).
We are interested in the case $\omega>2\Delta_{\mathbf{\check{p}}}$
and $k<\omega$. Since $a\ll1$ and $p\simeq p_{F}$, the second integral
in Eq. (\ref{hi}) may be dropped. By neglecting also the small contributions
from $a^{2}k^{2}\ll\epsilon_{p}^{2},\omega^{2}$ we get\[
\chi\left(K\right)=e^{2}\frac{a^{2}}{\omega^{2}}\int\ \allowbreak\frac{\Delta_{\mathbf{\check{p}}}^{2}}{\epsilon_{p}^{3}}\frac{d^{3}p}{\left(2\pi\right)^{3}}=\frac{1}{\omega^{2}}\frac{e^{2}n}{M^{\ast}}\int\frac{d\Omega_{\mathbf{p}}}{4\pi}\int_{0}^{\infty}\ \allowbreak\frac{\Delta_{\mathbf{\check{p}}}^{2}}{\epsilon_{p}^{3}}d\xi\]
 where $n$ is the number of baryons per unit volume. Since\begin{eqnarray*}
\int\frac{d\Omega_{\mathbf{p}}}{4\pi}\int_{0}^{\infty}\ \allowbreak\frac{\Delta_{\mathbf{\check{p}}}^{2}}{\epsilon^{3}}d\xi & = & \int\frac{d\Omega_{\mathbf{p}}}{4\pi}\int_{\Delta}^{\infty}\ \allowbreak\frac{\Delta_{\mathbf{\check{p}}}^{2}}{\epsilon^{2}}\frac{1}{\sqrt{\epsilon^{2}-\Delta_{\mathbf{\check{p}}}^{2}}}d\epsilon\\
 & = & \int\frac{d\Omega_{\mathbf{p}}}{4\pi}\int_{1}^{\infty}\ \allowbreak\frac{1}{x^{2}}\frac{1}{\sqrt{x^{2}-1}}dx=1\end{eqnarray*}
 we obtain \[
\chi\left(K\right)=\frac{\omega_{p}^{2}}{\omega^{2}},\ \ \ \ \omega_{p}^{2}=\frac{e^{2}n}{M^{\ast}}\]
 This agrees with the plasma frequency for a free gas of charged particles.

The energy exchange in the medium goes naturally as the temperature
scale. Therefore, the energy transferred to the radiated neutrino-pair
is $\omega\sim T\leq T_{c}$, while the plasma frequency $\omega_{p}$
is typically much larger than the critical temperature for Cooper
pairing. For instance, for a number density $n$ of the order of the
nuclear saturation density $n_{0}\simeq0.17$ fm and the effective
mass of the baryon $M^{\ast}$ of the order of the bare nucleon mass,
we obtain $\omega_{p}\sim10\ MeV$, while the critical temperature
for baryon pairing is about $1\ MeV$ or less. Under these conditions,
we obtain

\[
\tilde{\Gamma}^{0}\left(p-K,p\right)\simeq\frac{\omega^{2}}{\omega_{p}^{2}}\Gamma^{0}\left(p-K,p\right)\sim\frac{T_{c}^{2}}{\omega_{p}^{2}}\Gamma^{0}\left(p-K,p\right)\]
 Thus, in superconductors, the vector current contribution to the
neutrino radiation is suppressed additionally by a factor $\left(T_{c}^{2}/\omega_{p}^{2}\right)^{2}$:
this is the \textit{plasma screening effect}. The total suppression
factor, due to both the current conservation and the plasma effects,
is of the order \[
\left(T_{c}^{2}/\omega_{p}^{2}\right)^{2}V_{F}^{4}\lesssim10^{-6}.\]

\section{Energy losses in the axial channel}

The neutrino energy losses in the axial channel can also be obtained
with the aid of Fermi's golden rule. After integration over the phase
space of participating particles, this yields (see details in \cite{YKL98}):\begin{equation}
Q_{A}=C_{A}^{2}\frac{4G_{F}^{2}{p}_{F}M^{\ast}}{15\pi^{5}}\mathcal{N}_{\nu}T^{7}R_{A},\ \ \ R_{A}=\frac{C_{A}^{2}}{8\pi}\int d\Omega\
\int_{0}^{\infty}dx\;\frac{z^{6}dx}{\left(e^{z}+1\right)^{2}}I.\label{QA}\end{equation}
 Here $\mathcal{N}_{\nu}=3$ is the number of neutrino flavors, $z=\sqrt{x^{2}+\Delta_{\mathbf{\check{p}}}^{2}/T^{2}}$,
and the outer integration is performed over the orientations of the
nucleon momentum $\mathbf{p}$. The matrix elements of the axial current
\[
\mathbf{A}\left(\mathbf{\check{p}},\eta\right)\mathbf{\equiv}\left\langle \Psi_{-\mathbf{p,}-\eta^{\prime}}^{\dagger}\left\vert \mathbf{\hat{A}}\right\vert \Psi_{\mathbf{p,\eta}}\right\rangle \]
 come into this expression under the combination \begin{equation}
I=I_{xx}+I_{yy}+I_{zz}\label{I}\end{equation}
 with\[
I_{ii}=\sum_{\eta\eta^{\prime}}A_{i}\left(\mathbf{\check{p}},\eta\right)A_{i}^{\ast}\left(\mathbf{\check{p}},\eta^{\prime}\right)\]

The collective mode considered in previous sections represents density
oscillations of the condensate, and does not influence the axial weak
vertex of a quasi-particle, if the average spin projection of the
bound pair is zero. This happens in the singlet-state pairing and
in triplet pairing with total momentum projection $m_{j}=0$. In these
cases, the axial vertex may be taken in its bare form $A_{\mathrm{i}}^{bare}=\hat{\Sigma_{i}}$,
where the spin operator \begin{equation}
\hat{\Sigma_{i}}=\hat{\sigma_{i}}\hat{\tau}_{3}\equiv\begin{pmatrix}\hat{\sigma_{i}} & 0\\
0 & -\hat{\sigma_{i}}\end{pmatrix},\label{S}\end{equation}
 with $\hat{\sigma_{i}}=\left(\hat{\sigma}_{x},\hat{\sigma}_{y},\hat{\sigma}_{z}\right)$
being the Pauli spin matrices, is defined according to the relation\[
\hat{\Sigma}_{z}\Psi_{\mathbf{p},\eta}=\eta\Psi_{\mathbf{p},\eta},\]
 where \[
\Psi_{\mathbf{p},\eta}=\left(\begin{array}{c}
u_{\mathbf{p}}\chi_{\eta}\\
v_{\mathbf{p}}\chi_{-\eta}\end{array}\right)\]
 is the eigenstate with spin projection $s_{z}=\eta/2=\pm1/2$ along
the $Z$-axis.

A direct evaluation of the matrix elements of the bare axial vertex
(\ref{S}) for a quasi-particle transition into a hole by making use
of the wave functions (\ref{pvf}), (\ref{nvf}) gives $I=0$ in the
case of singlet pairing. For triplet pairing with $m_{j}=0$ we obtain\[
I_{xx}=I_{yy}=\frac{\Delta_{\mathbf{\check{p}}}^{2}}{\epsilon_{p}^{2}}\left(1-\cos2\varphi\right),\ \ \ I_{zz}=2\frac{\Delta_{\mathbf{\check{p}}}^{2}}{\epsilon_{p}^{2}}\]
 which, after averaging over the azimuthal angle $\varphi$ of the
quasi-particle momentum $\mathbf{p}$, gives \begin{equation}
I=4\frac{\Delta_{\mathbf{\check{p}}}^{2}}{\epsilon_{p}^{2}}.\label{I0}\end{equation}
 This confirms the corresponding result obtained in \cite{YKL98}.

In the case of a triplet-state pairing with maximal momentum projection
$\left\vert m_{j}\right\vert =2$, we obtain a different result. In
this case, the spin projection of the Cooper pair is conserved, and
collective density oscillations of the condensate are accompanied
by oscillations of the spin density, i.e. by the axial current. The
corresponding gap matrix of the quasi-particles has the form $\hat{D}=\Delta_{\mathbf{\check{p}}}\hat{\Upsilon}\left(\mathbf{\check{p}}\right)$
with \cite{Tamagaki}\[
\hat{\Upsilon}\left(\mathbf{\check{p}}\right)=\left(\begin{array}{cc}
-e^{i\varphi} & 0\\
0 & e^{-i\varphi}\end{array}\right)=-\hat{\sigma}_{z}\cos\varphi-i\sin\varphi.\]
 Therefore, the operator $\hat{\Sigma}_{z}$ commutes with the quasi-particle
Hamiltonian $H_{0}=\xi_{\mathbf{p}}\hat{\tau}_{3}+\hat{\Lambda}\left(\mathbf{\check{p}}\right)$,
where $\hat{\Lambda}$ is given by Eq. (\ref{Lam}). This modifies
the $z$-component of the axial vertex, which should be found from
the equation\[
\hat{A}_{z}\left(p-K,p\right)=\hat{\tau}_{3}\hat{\sigma}_{z}+i\int\frac{d^{4}p^{\prime}}{\left(2\pi\right)^{4}}\ \hat{\tau}_{3}\text{\c{G}}\left(p^{\prime}-K\right)\hat{A}_{z}\left(p^{\prime}-K,p^{\prime}\right)\text{\c{G}}\left(p^{\prime}\right)\hat{\tau}_{3}V_{pp^{\prime}}.\]
 After multiplication of this equation by $\hat{\sigma}_{z}$ from
the right-hand side, and introducing the new function \begin{equation}
\hat{B}\equiv\hat{A}_{z}\left(p-K,p\right)\hat{\sigma}_{z}\label{B}\end{equation}
 we arrive to the following equation \[
\hat{B}\left(p-K,p\right)=\hat{\tau}_{3}+i\int\frac{d^{4}p^{\prime}}{\left(2\pi\right)^{4}}\ \hat{\tau}_{3}\text{\c{G}}\left(p^{\prime}-K\right)\hat{B}\left(p^{\prime}-K,p^{\prime}\right)\hat{\sigma}_{z}\text{\c{G}}\left(p^{\prime}\right)\hat{\sigma}_{z}\hat{\tau}_{3}V_{pp^{\prime}}.\]
 By using the fact that $\hat{\sigma}_{z}$ commutes with \c{G}$\left(p^{\prime}\right)$,
one can reduce this equation to the form\[
\hat{B}\left(p-K,p\right)=\hat{\tau}_{3}+i\int\frac{d^{4}p^{\prime}}{\left(2\pi\right)^{4}}\ \hat{\tau}_{3}\text{\c{G}}\left(p^{\prime}-K\right)\hat{B}\left(p^{\prime}-K,p^{\prime}\right)\text{\c{G}}\left(p^{\prime}\right)\hat{\tau}_{3}V_{pp^{\prime}}\]
 which is identical to Eq. (\ref{Gam}) for $\mu=0$. Then, the solution
has a form analogous to Eq. (\ref{G0}). We obtain \[
\hat{B}\left(p-K,p\right)=\hat{\tau}_{3}-2\hat{\tau}_{3}\hat{\Lambda}\left(\mathbf{\check{p}}\right)\frac{\omega}{\omega^{2}-a^{2}k^{2}}.\]
 From Eq. (\ref{B}) we find\[
\hat{A}_{z}\left(p-K,p\right)=\hat{\tau}_{3}\hat{\sigma}_{z}-2\hat{\tau}_{3}\hat{\Lambda}\left(\mathbf{\check{p}}\right)\hat{\sigma}_{z}\frac{\omega}{\omega^{2}-a^{2}k^{2}},\]
 or, in an explicit form:\[
\hat{A}_{z}\left(p-K,p\right)=\left(\begin{array}{cc}
\hat{\sigma}_{z} & 0\\
0 & -\hat{\sigma}_{z}\end{array}\right)-2\left(\begin{array}{cc}
0 & \hat{\sigma}_{z}\hat{\Upsilon}\\
-\hat{\sigma}_{z}\hat{\Upsilon}^{\dagger} & 0\end{array}\right)\frac{\omega\Delta_{\mathbf{\check{p}}}}{\omega^{2}-a^{2}k^{2}}.\]
 The transverse components of the axial vertex are not modified\[
\hat{A}_{x}=\hat{\tau}_{3}\hat{\sigma}_{x}=\left(\begin{array}{cc}
\hat{\sigma}_{x} & 0\\
0 & -\hat{\sigma}_{x}\end{array}\right),\ \ \ \hat{A}_{y}=\hat{\tau}_{3}\hat{\sigma}_{y}=\left(\begin{array}{cc}
\hat{\sigma}_{y} & 0\\
0 & -\hat{\sigma}_{y}\end{array}\right).\]
 Direct evaluation of the matrix elements for $k\ll p_{F}$ gives,
in this case:

\[
A_{z}\left(\mathbf{\check{p}},\eta\right)\simeq-\eta\left(\frac{\Delta_{\mathbf{\check{p}}}}{\epsilon_{\mathbf{p}}}-\frac{2\Delta_{\mathbf{\check{p}}}\omega}{\omega^{2}-a^{2}k^{2}}\right)\delta_{-\eta^{\prime},\eta}\simeq\frac{1}{3}\eta V_{F}^{2}\frac{k^{2}}{\omega^{2}}\frac{\Delta_{\mathbf{\check{p}}}}{\epsilon_{p}}\delta_{-\eta^{\prime},\eta},\]
 \[
A_{x}\left(\mathbf{\check{p}},\eta\right)=-\frac{\Delta_{\mathbf{\check{p}}}}{2\epsilon_{p}}\left(1-e^{2\eta i\varphi}\right)\delta_{\eta,\eta^{\prime}},\]
 \[
A_{y}\left(\mathbf{\check{p}},\eta\right)=-\frac{\Delta_{\mathbf{\check{p}}}}{2\epsilon_{p}}i\left(1-e^{2\eta i\varphi}\right)\delta_{\eta,\eta^{\prime}}\left(\delta_{\uparrow,\eta}-\delta_{\downarrow,\eta}\right).\]
 Due to the contribution of the collective mode, the matrix element
$A_{z}$ is $V_{F}^{2}$ times smaller than $A_{x}$ and $A_{y}$
and, thus, can be neglected. After averaging over the azimuthal angle
$\varphi$ of the quasi-particle momentum, we obtain \begin{equation}
I\simeq I_{xx}+I_{yy}=2\frac{\Delta_{\mathbf{\check{p}}}^{2}}{\epsilon_{p}^{2}}.\label{I2}\end{equation}

Thus, in the case of $^{3}P_{2}$ pairing with total momentum projection
$\left\vert m_{j}\right\vert =2$, the value of $I$, given by Eq.
(\ref{I2}), is twice smaller, and the neutrino energy losses are
proportionally suppressed.

\section{Discussion and conclusions}

We have considered the problem of conservation of the vector weak
current in the theory of neutrino-pair radiation from Cooper pairing
in neutron stars. The correction to the vector weak vertex is calculated
within the same order of approximation as the quasi-particle propagator
is modified by the pairing interaction in the system. This correction
restores the conservation of the vector weak current in the quasi-particle
transition into the paired state. As a result, in the nonrelativistic
baryon system, the matrix element of the vector current is $V_{F}^{2}$
times smaller than previous estimations. This means that the vector
weak current contribution to neutrino radiation caused by Cooper paring
is $V_{F}^{4}$ times smaller than it was thought before. The vector
weak current contribution from pairing of charged baryons is suppressed
additionally by a factor $\sim\left(T_{c}^{2}/\omega_{p}^{2}\right)^{2}$
due to plasma screening. The total suppression factor due to both
the current conservation and the plasma effects is of the order \[
\left(T_{c}^{2}/\omega_{p}^{2}\right)^{2}V_{F}^{4}\lesssim10^{-6}.\]

In papers \cite{L00}, \cite{L01}, a special case of proton pairing
has been considered which accounts for electron-proton correlations,
resulting in an increasing of the neutrino emissivity. This, however,
takes place only because of a very small weak vector coupling constant
of the proton with respect to that for the electron. In our calculations,
incorporation of the electron-proton correlations would result in
the replacement of the vector weak coupling constant of the proton
by a form-factor, which is proportional (to the leading order) to
the electron coupling constant. This could enlarge the effective proton
vector weak coupling to almost the same order of magnitude as that
for neutrons. However, the cancellation of the temporal component
of the transition current due to current conservation is much more
stronger, and makes negligible the neutrino emission from the proton
pairing even with taking into account the electron-proton correlations.
Thus, the neutrino energy losses due to singlet-state pairing of baryons
can, in practice, be neglected in simulations of neutron star cooling.
This makes unimportant the neutrino radiation from pairing of protons
or hyperons.

The neutrino radiation from triplet pairing occurs through the axial
weak currents. In the case of $^{3}P_{2}$ pairing with total momentum
projection $m_{j}=0$, the corresponding neutrino emissivity is given
by \begin{equation}
Q\left(m_{j}=0\right)=C_{A}^{2}\frac{8G_{F}^{2}{p}_{F}M^{\ast}}{15\pi^{5}}\mathcal{N}_{\nu}T^{7}F_{t},\label{Q0}\end{equation}
 with\begin{equation}
F_{t}=\frac{1}{4\pi}\int d\Omega\ y^{2}\int_{0}^{\infty}dx\;\frac{\left(x^{2}+y^{2}\right)^{2}}{\left(e^{\sqrt{x^{2}+y^{2}}}+1\right)^{2}}.\label{F}\end{equation}
 This formula can be obtained directly from the corresponding expression
suggested in \cite{YKL98} by omitting the contribution proportional
to the vector weak coupling constant $C_{V}^{2}$. As a result, the
corresponding neutrino energy losses are suppressed by about 25\%\[
\frac{Q\left(m_{j}=0\right)}{Q_{\mathrm{YKL}}\left(m_{j}=0\right)}=\frac{2C_{A}^{2}}{C_{V}^{2}+2C_{A}^{2}}=\frac{3.17}{4.17}\]
 with respect to those calculated in \cite{YKL98}.

In the case of $^{3}P_{2}$ pairing with total momentum projection
$\left\vert m_{j}\right\vert =2$, the neutrino energy losses in the
axial channel are additionally suppressed due to the collective contribution
of the spin-density fluctuations in the condensate: \begin{equation}
Q\left(\left\vert m_{j}\right\vert =2\right)=C_{A}^{2}\frac{4G_{F}^{2}{p}_{F}M^{\ast}}{15\pi^{5}}\mathcal{N}_{\nu}T^{7}F_{t}.\label{Q2}\end{equation}
 Together with the vanishing of the vector contribution, the total
suppression of the neutrino emissivity with respect to that obtained
in \cite{YKL98} is about 3 times\[
\frac{Q\left(\left\vert m_{j}\right\vert =2\right)}{Q_{\mathrm{YKL}}\left(\left\vert m_{j}\right\vert =2\right)}=\frac{C_{A}^{2}}{C_{V}^{2}+2C_{A}^{2}}=\frac{1.58}{4.17}.\]

In this paper, we have considered the neutrino radiation from Cooper
pairing in nonrelativistic baryon matter. It is clear from the above
consideration, however, that conservation of the vector weak current
has to be restored also in the theory of neutrino radiation from pairing
of relativistic quarks \cite{prak}. This will be done in a future
work.

\selectlanguage{american} \textbf{Acknowledgments}

\selectlanguage{english} This work has been supported by Spanish
Grants AYA2004-08067-C01, FPA2005-00711 and GV2005-264.

{10}

\end{document}